\documentclass[twocolumn,aps,superscriptaddress,showpacs,nofootinbib,floatfix]{revtex4}
\usepackage{epsfig,bm,feynmf}
\usepackage{graphics}
\usepackage{amsmath}
\usepackage[normalem]{ulem}  
\usepackage[dvips]{color} 

\renewcommand\sout{\bgroup \color{red} \ULdepth=-.5ex \ULset}

\begin{document}


\title{Charmonia formation in quark-gluon plasma}


\author{Taesoo Song}\email{song@fias.uni-frankfurt.de}
\affiliation{Frankfurt Institute for Advanced Studies and Institute for Theoretical Physics, Johann Wolfgang Goethe Universit\"{a}t, Frankfurt am Main, Germany}


\begin{abstract}
Using the color evaporation model, the cross section for charmonium production in p+p collision is calculated in quark-gluon plasma.
The threshold energy for open charms is given by the free energy potential from lattice calculations, the initial charm quark pairs by the Pythia simulations, and their time evolution by solving the Langevin equation.
It is found that the threshold energy which decreases with temperature reduces the cross section while the invariant mass of charm pair which decreases by collisions enhances it.
As a result, charmonia production is suppressed by 30$\sim$50 \% while $J/\psi$ production is similar or enhanced compared to in vacuum.
\end{abstract}


\maketitle

\section{introduction}

Heavy quarkonium such as $J/\psi$ and $\Upsilon$ is an interesting probe to investigate the properties of the hot dense nuclear matter created by relativistic heavy-ion collisions.
It originates from the idea that the suppression of quarkonium in relativistic heavy-ion collisions is the signature of quark-gluon plasma (QGP) formation~\cite{Matsui:1986dk}, and such a suppression has been measured in many experiments~\cite{Alessandro:2004ap,Adare:2006ns,:2010px,Chatrchyan:2011pe,Abelev:2012rv,Manceau:2012ka}.
Now the modification of quakonium yield is understood as the result of many different kinds of nuclear matter effects.
They are classified into the cold and hot nuclear matter effects.
The examples of the former effect are the nuclear (anti-)shadowing, the Cronin effect, the nuclear absorption, and so on.
They mostly happen before quarkonium formation.
The latters are thermal decay and regeneration both in QGP and in hadron gas~\cite{Vogt:1999cu,Zhang:2000nc,Zhang:2002ug,Grandchamp:2002wp,Yan:2006ve,Linnyk:2008hp}, which take place after the quarkonium formation.

Quarknoium is formed from the heavy quark pair produced in parton-parton hard collisions.
The formation time of quarkonium is not short compared to the time for heavy quark pair production.
It ranges from a few tenths to a couple of fermis, depending on model~\cite{Karsch:1987zw,Blaizot:1988ec,Kharzeev:1999bh}.
Recently, the quarkonium formation time in QGP was calculated by using dispersion relations and heavy quark potential energies extracted from the Lattice Quantum Chromodynamics (LQCD)~\cite{Song:2013lov}.
It was found that the formation time increases with temperature and diverges near quarkonium dissociation temperature.
It is reasonable because the size of quarkonium increases with temperature and the formation must take longer time.

The temperature of the produced nuclear matter in relativistic heavy-ion collisions at the top energy of the Relativistic Heavy Ion Collider (RHIC) or the Large Hadron Collider (LHC) is much higher than the critical temperature for QGP phase transition and the QGP phase lasts for several fermis.
Therefore, it is necessary to consider the nuclear matter effect on heavy quark pair before it forms quarkonium in order to understand experimental data.

The color evaporation model is a simple but successful method to calculate the cross section for quarkonium production~\cite{Amundson:1996qr}.
It factorizes the initial production of heavy quark pair which is calculated in perturbative Quantum Chromodynamics (pQCD) and the quarkonium formation from the pair which is nonperturbative.
The latter is simplified such that a constant fraction of heavy quark pairs of which invariant mass is below continuum threshold forms a certain kind of quarkonium regardless of collision energy.

In this study, we apply the color evaporation model to the quarkonium formation in QGP.
Two modifications are made for this purpose:
In the model the continuum threshold energy for open charms is twice D meson mass in vacuum.
Since D meson is supposed to be dissolved in QGP, we substitute the threshold energy by the sum of dressed charm and anticharm quark masses which is obtained by separating them infinitely in LQCD.
The second modification is the heavy quark and heavy antiquark momentum distributions in QGP.
The latter is not negligible if quarkonium formation time is long.

This paper is organized as follows: In Sec.~\ref{CEM}, we briefly review the color evaporation model and modify it in QGP. We then describe charm and anticharm quark momentum distributions in QGP by using the Pythia event generator and the Langevin equation in Sec.~\ref{Langevin}. Finally results and a summary are given in Sec.~\ref{results} and \ref{summary} respectively.

\section{color evaporation model}\label{CEM}

Quarkonium is produced in nucleon-nucleon collisions through a heavy quark pair production.
Because the production of heavy quark pair requires large energy-momentum transfer, it is calculated in pQCD as following:
\begin{eqnarray}
\frac{d\sigma_{NN\rightarrow Q\bar{Q}}}{dM}(M)=\sum_{i,j=q,\bar{q},g}\int dx_1 dx_2 f_i(x_1;Q)f_j(x_2;Q)\nonumber\\
\times\frac{d\sigma_{ij\rightarrow Q\bar{Q}}}{dM}(M;Q),~~~~~~~
\end{eqnarray}
where $M$ is the invariant mass of heavy quark pair; $x_1$ and $x_2$ are the longitudinal momentum fractions of parton $i$ and $j$, which produce a heavy quark pair, in parton distribution functions $f_i$ and $f_j$.
The parton distribution functions and the differential cross section for heavy quark production from partons both depend on the scale $Q$, while the differential cross section from nucleon scattering, which is a physical quantity, does not.

Once heavy quark pair produced, color evaporates from it to be a color-singlet state by emitting or absorbing soft gluons.
Through this nonperturbative process, some pairs form bound states and the others turn to open heavy flavors.
In the color evaporation model the cross sections for hidden and the open charm productions are respectively estimated by~\cite{Amundson:1996qr}
\begin{eqnarray}
\sigma_{\rm hidden}=\frac{1}{9}\int_{2m_c}^{2m_D} dM \frac{d\sigma_{c\bar{c}}}{dM},\label{hidden}
\end{eqnarray}
and
\begin{eqnarray}
\sigma_{\rm open}=\frac{8}{9}\int_{2m_c}^{2m_D} dM \frac{d\sigma_{c\bar{c}}}{dM}
+\int_{2m_D} dM \frac{d\sigma_{c\bar{c}}}{dM},\label{open}
\end{eqnarray}
where $m_c$ is the bare charm quark mass and $m_D$ the D meson mass. The prefactors $1/9$ and $8/9$ are statistical probabilities for heavy quark pair to be a color singlet and a color octet respectively.

Because it is not easy to measure all bound states in experiment, an additional constant factor is multiplied to compare the model with experimental data:
\begin{eqnarray}
\sigma_{J/\psi}=\rho_{J/\psi}\sigma_{\rm hidden}.
\end{eqnarray}

The constant $\rho_{J/\psi}$ has a universal value regardless of collision energy whether in photoproduction or in hadronproduction~\cite{Amundson:1996qr}.

In this study, we apply the same evaporation model to the quarkonium production in QGP.
Because D meson is supposed to be dissolved in QGP, the $2m_D$ in Eq.~(\ref{hidden}) is substituted by the sum of the dressed charm and anticharm quark masses which is defined as $2m_c^*\equiv 2m_c+V(r=\infty,T)$, where $V(r,T)$ is the potential energy between charm and anticharm quarks at temperature $T$ in QGP, and $V(r=\infty,T)$ is the energy required to separate them infinitely.
It is consistent with Eq.~(\ref{hidden}), because $2m_D\simeq 2m_c+V(r=\infty,T=0)$ in potential model.
Therefore, Eq.~(\ref{hidden}) is generalized into

\begin{eqnarray}
\sigma_{\rm hidden}=\frac{1}{9}\int_{2m_c}^{2m_c^*} dM \frac{d\sigma_{c\bar{c}}}{dM}.\label{hidden2}
\end{eqnarray}

\begin{figure}[h]
\centerline{
\includegraphics[width=9 cm]{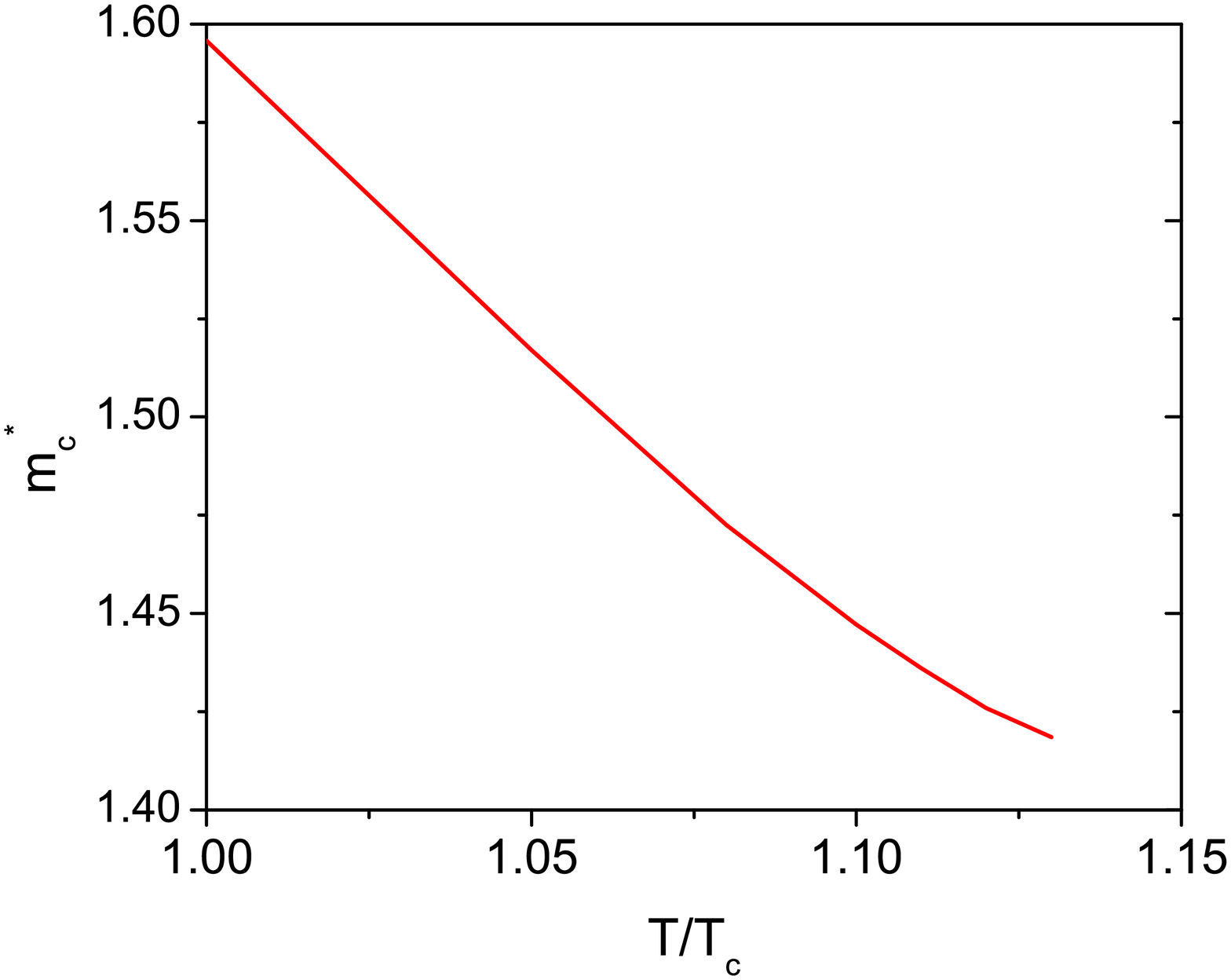}}
\caption{(Color online) Dressed charm quark mass from the free energy potential in LQCD~\cite{oai:arXiv.org:hep-ph/0505193,Kaczmarek:1900zz} as a function of temperature.}
\label{cmass}
\end{figure}

Figure \ref{cmass} shows the dressed charm quark mass from the free energy potential in LQCD~\cite{oai:arXiv.org:hep-ph/0505193,Kaczmarek:1900zz} as a function of temperature. The bare charm quark mass and the critical temperature $(T_c)$ are taken to be 1.25 GeV~\cite{Satz:2005hx} and 170 MeV respectively.
Solving the Schrodinger equation with the free energy potential, no bound state is found above 1.13 $T_c$.
As temperature increases, the binding of charmonium becomes weak and the dressed charm quark mass decreases.
The figure implies that the window for charmonium production becomes narrow in Eq.~(\ref{hidden2}) as temperature increases, and it will suppress charmonium production at high temperature.

\section{heavy quark distributions in QGP}\label{Langevin}

While heavy quark pair is produced promptly by hard collision, it takes time for the pair to form a quarkonium.
The time varies from several tenths to a couple of fermis, depending on model~\cite{Karsch:1987zw,Blaizot:1988ec,Kharzeev:1999bh}.
Recently the formation time of quarkonium in QGP is calculated by using dispersion relations and heavy quark potential energies extracted from LQCD~\cite{Song:2013lov}.
It was found that the formation time increases with temperature and diverges near the dissociation temperature of quarkonium.

\begin{figure}[h]
\centerline{
\includegraphics[width=9 cm]{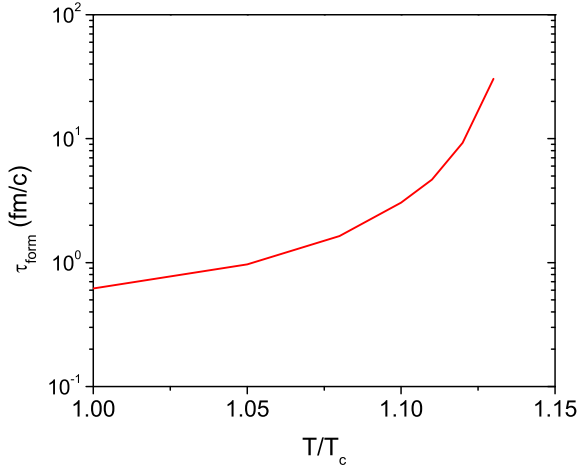}}
\caption{(Color online) The formation time of $J/\psi$ in QGP as a function of temperature with the free energy from LQCD being taken for the potential energy.}
\label{tform}
\end{figure}

Figure \ref{tform} shows the formation time of $J/\psi$ in QGP with the free energy from LQCD being taken for the potential energy between charm quark pair.
The formation time is about 0.6 fm/c at $T_c$ and then increases up to 30 fm/c near the dissociation temperature of $J/\psi$.
If the formation time is long, the energy-momenta of heavy quark and heavy antiquark will be modified from their initial ones.
Then Eq.~(\ref{hidden2}) is changed into
\begin{eqnarray}
\sigma_{\rm hidden}=\frac{1}{9}\int_{2m_c}^{2m_c^*} dM f_{c\bar{c}}(M,\tau=\tau_{\rm form}),\label{hidden3}
\end{eqnarray}
where $f_{c\bar{c}}(M,\tau=\tau_{\rm form})$ is the invariant mass distribution function of charm quark pairs at charmonium formation time.

We use the Langevin equation to get the time evolution of invariant mass distribution.
The macroscopic Langevin equation reads:
\begin{eqnarray}
\frac{dp_i}{dt}=\xi_i(t) -\eta_D p_i,\label{langevin}
\end{eqnarray}
where $\xi_i(t)$ and $\eta_D$ are random momentum kicks and momentum drag coefficient respectively.
The random momentum kicks have the correlation~\cite{Moore:2004tg},
\begin{eqnarray}
\langle \xi_i(t)\xi_j(t^\prime)\rangle=\kappa~\delta_{ij}\delta(t-t^\prime),\label{corr1}
\end{eqnarray}
where $3\kappa$ is the mean-squared momentum transfer per unit time.

From the solution of the Langevin equation, Eq.~(\ref{langevin}),
\begin{eqnarray}
p_i(t)=\int_{-\infty}^t dt^\prime e^{-\eta_D(t-t^\prime)}\xi_i(t^\prime),
\end{eqnarray}
derived are the relation~\cite{Moore:2004tg},
\begin{eqnarray}
3m_cT=\langle {\bf p}^2\rangle=\sum_i\langle p_i(0)p_i(0)\rangle~~~~~~~~~~~~~~~~~~~~~~~\nonumber\\
=\sum_i\int_{-\infty}^0 dt_1dt_2 e^{\eta_D(t_1+t_2)}\langle\xi_i(t_1)\xi_i(t_2)\rangle=\frac{3\kappa}{2\eta_D}
\label{temperature}
\end{eqnarray}
and the diffusion constant in space~\cite{Moore:2004tg}, which is denoted by $D$, for a particle starting at $(t,\vec{x})=(0,\vec{0})$,
\begin{eqnarray}
6Dt=\sum_i\langle x_i(t)x_i(t)\rangle~~~~~~~~~~~~~~~~~~~~~~~~~~~~~~~~~\nonumber\\
=\sum_i\int_0^tdt_1\int_0^tdt_2\bigg\langle\frac{p_i(t)}{m_c}\frac{p_j(t)}{m_c}\bigg\rangle
=\frac{3\kappa}{m_c^2\eta_D^2}t,
\label{diffusion}
\end{eqnarray}
assuming $t\eta_D\gg 1$.

Considering nonzero initial momentum, the random momentum kick is separated into longitudinal and transverse components depending on the direction of heavy quark motion, $\xi^i=\xi^i_L+\xi^i_T$.
Each component has the following correlations:
\begin{eqnarray}
\langle \xi^i_L(t)\xi^j_L(t^\prime)\rangle=\kappa_L\hat{p}^i\hat{p}^j\delta(t-t^\prime),~~~~~~~~~~\nonumber\\
\langle \xi^i_T(t)\xi^j_T(t^\prime)\rangle=\kappa_T(\delta^{ij}-\hat{p}^i\hat{p}^j)\delta(t-t^\prime),\nonumber\\
\langle \xi^i_T(t)\xi^j_L(t^\prime)\rangle=0,~~~~~~~~~~~~~~~~~~~~~~~~~~~
\label{corr2}
\end{eqnarray}
where $\hat{p}_i$ is the unit vector of heavy quark momentum.
In general $\kappa_L$ and $\kappa_T$ are functions of heavy quark momentum and temperature.
As heavy quark momentum decreases, $\kappa_L$ and $\kappa_T$ close to each other and Eq.~(\ref{corr2}) returns to Eq.~(\ref{corr1}).

The solution of the Langevin equation with nonzero initial momentum is given by
\begin{eqnarray}
p_i(t)=\int_{t_0}^t dt^\prime e^{-\eta_D(t-t^\prime)}\xi_i(t^\prime)+p_i(t_0)~e^{-\eta_D(t-t_0)},
\end{eqnarray}
assuming $\eta_D$ does not depend on heavy quark momentum, and the expectation value of squared momentum as a function of time by
\begin{eqnarray}
\langle {\bf p}^2(t)\rangle=\int_{t_0}^t dt^\prime \{\kappa_L(t^\prime)+2\kappa_T(t^\prime)\}~e^{-2\eta_D(t-t^\prime)}\nonumber\\
+{\bf p}^2(t_0)~e^{-2\eta_D(t-t_0)}.~~~~~~~~~~
\label{psquare}
\end{eqnarray}
We note that there is not a mixed term in Eq.~(\ref{psquare}) due to no correlation between initial heavy quark momentum and random momentum kick.
The first term in Eq.~(\ref{psquare}) is attributed to momentum diffusion of heavy quark and the second term to the attenuation of initial momentum.

We assume that later heavy quark has a gaussian momentum distribution as
\begin{eqnarray}
f_c(\vec{p})=\frac{1}{(2\pi)^{3/2}\sigma_L\sigma_T^2}\exp\bigg[-\frac{\{p_L-p_c(t)\}^2}{2\sigma_L^2}-\frac{p_T^2}{2\sigma_T^2}\bigg],
\label{gaussian}
\end{eqnarray}
which is centered at
\begin{eqnarray}
p_c(t)=p(t_0)~e^{-\eta_D(t-t_0)}
\end{eqnarray}
with the longitudinal and transverse widths being respectively
\begin{eqnarray}
\sigma_L^2=\int_{t_0}^t dt^\prime \kappa_L(t^\prime)~e^{-2\eta_D(t-t^\prime)},~~\nonumber\\
\sigma_T^2=2\int_{t_0}^t dt^\prime \kappa_T(t^\prime)~e^{-2\eta_D(t-t^\prime)}.
\label{widths}
\end{eqnarray}

For numerical calculations, $2\pi TD$ is taken to be 2 from LQCD~\cite{Ding:2011hr}. 
Once the diffusion constant, $D$, is given, $\eta_D$, $\kappa_L$ and $\kappa_T$ for static charm quark are obtained from Eq.~(\ref{temperature}) and (\ref{diffusion}):
\begin{eqnarray}
\eta_D=\frac{T}{m_cD},~~~\kappa_L=\kappa_T=\frac{2T^2}{D}.
\end{eqnarray}
And then we assume $\kappa_L$ and $\kappa_T$ have the momentum dependence derived in pQCD~\cite{Moore:2004tg} and $\eta_D$ is a function only of temperature.
The latter assumption is supported also by pQCD calculations~\cite{Moore:2004tg}.
A sufficient time later $p_c(t)$ disappears, and Eq.~(\ref{gaussian}) turns to a thermal distribution at the temperature given by Eq.~(\ref{temperature}).

\begin{figure}[h]
\centerline{
\includegraphics[width=9 cm]{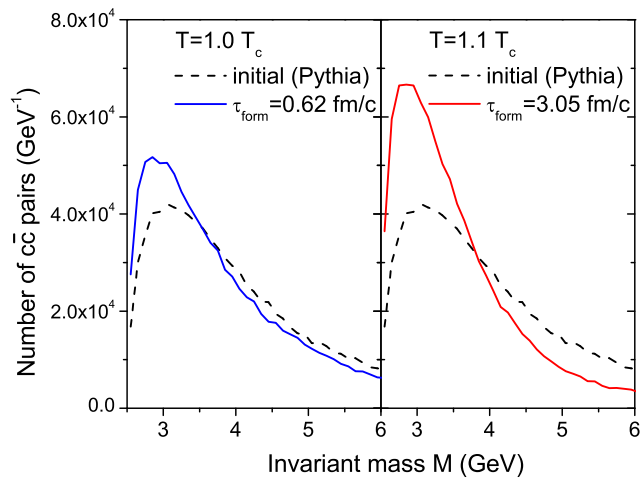}}
\caption{(Color online) Distributions of charm quark pairs as functions of invariant mass $M$ at initial productions and at charmonium formation times for temperatures of 1.0 $T_c$ and 1.1 $T_c$. The initial distribution is given at $\sqrt{s}=$ 200 GeV in p+p collisions by the Pythia simulations~\cite{Sjostrand:2006za} and $2\pi TD$ is taken to be 2 from LQCD~\cite{Ding:2011hr}.}
\label{distributions}
\end{figure}

In our study, initial charm and anticharm quarks are given by the Pythia simulations~\cite{Sjostrand:2006za}.
The bare charm quark mass, $m_c$, is tuned to be 1.25 GeV for consistency.
The black dashed lines in figure \ref{distributions} are the initial distributions of charm quark pairs as functions of invariant mass.
For each temperature one hundred thousand charm quark pairs are generated.

Then Eq.~(\ref{widths}) is calculated by multiplying $e^{-2\eta_D \Delta t}$ and adding $\kappa_L(t)\Delta t$ and $2\kappa_T(t)\Delta t$, respectively, to previous longitudinal and transverse widths at each time step till the charmonium formation time:
\begin{eqnarray}
\sigma_L^2(t_{n+1})= \sigma_L^2(t_{n})~e^{-2\eta_D\Delta t}+\kappa_L(t_n)\Delta t,~~\nonumber\\
\sigma_T^2(t_{n+1})=\sigma_T^2(t_{n})~e^{-2\eta_D\Delta t}+2\kappa_T(t_n)\Delta t,
\end{eqnarray}
where $\kappa_L(t_n)$ and $\kappa_T(t_n)$ are functions of charm quark momentum which is determined from Eq.~(\ref{psquare}) by
\begin{eqnarray}
\langle {\bf p}^2(t_n)\rangle=\sigma_L^2(t_{n})+\sigma_T^2(t_{n})+{\bf p}^2(t_0)~e^{-2\eta_D(t_n-t_0)}.
\end{eqnarray}
The same steps are taken for anticharm quark.

The formation time of charmonium is calculated in the rest frame of charm quark pair:
\begin{eqnarray}
\tau=\int dt\sqrt{1-v^2(t)},
\end{eqnarray}
where $v(t)$ is the velocity of charm quark pair.
So obtained $\sigma_L^2$ and $\sigma_T^2$ at the formation time are substituted into Eq.~(\ref{gaussian}) and the momenta of charm and anticharm quarks are decided by the Monte Carlo method.

The solid lines in figure \ref{distributions} are the distributions of charm quark pairs as functions of invariant mass at charmonium formation times for temperatures of 1.0 $T_c$ and 1.1 $T_c$. Because the formation time is much longer at 1.1 $T_c$ than at 1.0 $T_c$, the distribution is more shifted to lower invariant mass.
In the view of the color evaporation model, it enhances charmonium production at high temperature in QGP, ignoring the change of dressed charm quark mass which was discussed in the previous section.

\section{results}\label{results}

Now we combine Sec.~\ref{CEM} and \ref{Langevin} to study the nuclear matter effect on charmonium formation in QGP.

\begin{figure}[h]
\centerline{
\includegraphics[width=9 cm]{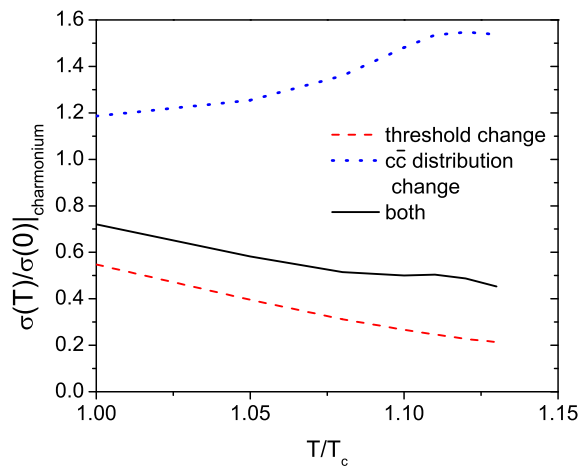}}
\caption{(Color online) The ratios of charmonium production cross section in QGP to that in vacuum as functions of temperature at $\sqrt{s}=$ 200 GeV in p+p collision. The dashed line is the ratio from the threshold energy change, the dotted line from the invariant mass change, and the solid line from both.}
\label{raaf}
\end{figure}

Figure \ref{raaf} shows the ratios of charmonium production cross section in QGP to that in vaccum as functions of temperature at $\sqrt{s}=$ 200 GeV in p+p collision.
The dashed line is the ratio of cross sections from the threshold energy change.
The reduced masses of dressed charm and anticharm quarks suppress charmonium production by lowering the threshold energy for open charm production.
The dotted line is the ratio from the invariant mass change of charm quark pair.
The charm and anticharm spectra softened in QGP increase the cross section.
The ratio begins to saturate above 1.1 $T_c$, where the formation time of charmonium is long enough to thermalize charm quarks.
The solid line is the ratio from both effects.
It is shown that the effect of threshold energy change is stronger than that of invariant mass change.
The cross section ratio decreases to 70 \% at 1.0 $T_c$ and 50 \% at 1.1 $T_c$.

However, it does not mean the $J/\psi$ suppression which is often measured in experiments.
Among the charmonia produced at $\sqrt{s}=$ 200 GeV in p+p collisions, roughly 50 \% of them are $J/\psi$ and others excited states such as $\chi_c$ and $\psi^\prime$~\cite{Song:2011xi}.
Taking the free energy from LQCD for the potential energy between charm and anticharm quarks, the excited states of charmonium disappear above $T_c$~\cite{Satz:2005hx}.
Therefore the charmonium in figure \ref{raaf} is always $J/\psi$ and the ratio of $J/\psi$ production cross sections would be twice, which is between 1.0 and 1.4.

Furthermore, it is known that about 40 \% of $J/\psi$ are produced through the decay of $\chi_c$ and $\psi^\prime$~\cite{Song:2011xi}.
Subtracting this contribution, the cross section ratio for $J/\psi$ is supposed to be 0.6 in QGP.
In comparison with it our results are much larger.

For the potential energy between heavy quark pair, the internal energy from LQCD can be used as well~\cite{Satz:2005hx,Kaczmarek:1900zz}.
In this case the charmonium production will be more enhanced, because the stronger binding makes the formation time shorter and the threshold energy for open charms higher.
However, the internal energy potential from LQCD has a overshooting problem near $T_c$ where the $J/\psi$ mass and the binding energy are higher than those in vacuum~\cite{Dumitru:2009ni}. And a recent study using the QCD sum rule supports the free energy potential rather than the internal energy one~\cite{Lee:2013dca}.

\section{summary}\label{summary}

We studied the nuclear matter effect on charomium formation in QGP by using the color evaporation model.
The color evaporation model was generalized to QGP phase by substituting the dressed charm and anticharm quark masses for the threshold energy for open charms and by modifying the invariant mass distribution of charm quark pairs in QGP.
The former was done by using the free energy from lattice calculations as the potential energy between charm and anticharm quarks.
The dressed charm and anticharm quark masses decrease with temperature and it suppresses charmonium production in QGP.
The latter was carried out by using the Langevin equation.
Initial charm and anticharm quarks in p+p collisions are generated by the Pythia simulations.
The drag and diffusion coefficients for a static heavy quark are obtained from lattice calculations and their momentum dependence from pQCD results.
The nuclear matter softens charm and anticharm spectra in QGP, which enhances charmonium production in QGP.

Combining both effects, we found that the cross section for charmonium production in p+p collision is reduced by 30 to 50 \% in QGP compared to in vacuum.
However, since $J/\psi$ is the only charmonium formed in QGP for the lattice free energy potential, the above results are interpreted as that $J/\psi$ production is similar or enhanced in QGP.

\section*{Acknowledgements}

This work was supported by the DFG.


\begin{thebibliography}{99}
\bibitem{Matsui:1986dk}
  T.~Matsui and H.~Satz,
  Phys.\ Lett.\  B {\bf 178}, 416 (1986).

\bibitem{Alessandro:2004ap}
  B.~Alessandro {\it et al.}  [NA50 Collaboration],
  Eur.\ Phys.\ J.\  C {\bf 39}, 335 (2005).

\bibitem{Adare:2006ns}
  A.~Adare {\it et al.}  [PHENIX Collaboration],
  Phys.\ Rev.\ Lett.\  {\bf 98}, 232301 (2007).

\bibitem{:2010px}
  G.~Aad {\it et al.}  [Atlas Collaboration],
  Phys.\ Lett.\  B {\bf 697}, 294 (2011).

\bibitem{Chatrchyan:2011pe}
  S.~Chatrchyan {\it et al.}  [CMS Collaboration],
  Phys.\ Rev.\ Lett.\  {\bf 107}, 052302 (2011).

\bibitem{Abelev:2012rv}
  B.~Abelev {\it et al.}  [ALICE Collaboration],
  Phys.\ Rev.\ Lett.\  {\bf 109}, 072301 (2012).

\bibitem{Manceau:2012ka}
  L.~Manceau [ALICE Collaboration],
  Nucl.\ Phys.\ Proc.\ Suppl.\  {\bf 234}, 321 (2013).

\bibitem{Vogt:1999cu}
  R.~Vogt,
  Phys.\ Rept.\  {\bf 310}, 197-260 (1999).

\bibitem{Zhang:2000nc}
  B.~Zhang, C.~M.~Ko, B.~A.~Li, Z.~w.~Lin and B.~H.~Sa,
  Phys.\ Rev.\  C {\bf 62}, 054905 (2000)

\bibitem{Zhang:2002ug}
  B.~Zhang, C.~M.~Ko, B.~A.~Li, Z.~W.~Lin and S.~Pal,
  Phys.\ Rev.\  C {\bf 65}, 054909 (2002)

\bibitem{Grandchamp:2002wp}
  L.~Grandchamp and R.~Rapp,
  Nucl.\ Phys.\  A {\bf 709}, 415 (2002).

\bibitem{Yan:2006ve}
  L.~Yan, P.~Zhuang and N.~Xu,
  Phys.\ Rev.\ Lett.\  {\bf 97}, 232301 (2006).

\bibitem{Linnyk:2008hp}
  O.~Linnyk, E.~L.~Bratkovskaya and W.~Cassing,
  Int.\ J.\ Mod.\ Phys.\ E {\bf 17}, 1367 (2008).

\bibitem{Karsch:1987zw}
  F.~Karsch, R.~Petronzio,
  Z.\ Phys.\ C {\bf 37}, 627 (1988).

\bibitem{Blaizot:1988ec}
  J.~P.~Blaizot, J.~-Y.~Ollitrault,
  Phys.\ Rev.\ D  {\bf 39}, 232 (1989).

\bibitem{Kharzeev:1999bh}
  D.~Kharzeev and R.~L.~Thews,
  Phys.\ Rev.\ C {\bf 60} (1999) 041901.

\bibitem{Song:2013lov}
  T.~Song, C.~M.~Ko and S.~H.~Lee,
  Phys.\ Rev.\ C {\bf 87}, 034910 (2013).

\bibitem{Amundson:1996qr}
  J.~F.~Amundson, O.~J.~P.~Eboli, E.~M.~Gregores and F.~Halzen,
  Phys.\ Lett.\ B {\bf 390}, 323 (1997).

\bibitem{oai:arXiv.org:hep-ph/0505193}
  S.~Digal, O.~Kaczmarek, F.~Karsch and H.~Satz,
  Eur.\ Phys.\ J.\ C {\bf 43}, 71 (2005).

\bibitem{Kaczmarek:1900zz}
  O.~Kaczmarek,
  Eur.\ Phys.\ J.\ C {\bf 61}, 811 (2009).

\bibitem{Satz:2005hx}
  H.~Satz,
  J.\ Phys.\ G {\bf 32}, R25 (2006).

\bibitem{Moore:2004tg}
  G.~D.~Moore and D.~Teaney,
  Phys.\ Rev.\ C {\bf 71}, 064904 (2005).

\bibitem{Ding:2011hr}
  H.~T.~Ding, A.~Francis, O.~Kaczmarek, F.~Karsch, H.~Satz and W.~Soldner,
  J.\ Phys.\ G {\bf 38}, 124070 (2011).

\bibitem{Sjostrand:2006za}
  T.~Sjostrand, S.~Mrenna and P.~Z.~Skands,
  JHEP {\bf 0605}, 026 (2006).

\bibitem{Song:2011xi}
  T.~Song, K.~C.~Han and C.~M.~Ko,
  Phys.\ Rev.\ C {\bf 84}, 034907 (2011).

\bibitem{Dumitru:2009ni}
  A.~Dumitru, Y.~Guo, A.~Mocsy and M.~Strickland,
  Phys.\ Rev.\ D {\bf 79}, 054019 (2009).

\bibitem{Lee:2013dca}
  S.~H.~Lee, K.~Morita, T.~Song and C.~M.~Ko,
  arXiv:1304.4092 [nucl-th].

\end{thebibliography}
\end{document}